\documentclass[12pt]{article}

 at 10truept

\def\rdots{\mathinner{\mkern1mu\raise1pt\vbox{\kern1pt\hbox{.}}\mkern2mu
   \raise4pt\hbox{.}\mkern2mu\raise7pt\hbox{.}\mkern1mu}}
\newcommand{\ve}{\varepsilon}
\newcommand{\la}{\lambda}
\newcommand{\ra}{\rightarrow}
\newcommand{\Z}{{\rm Z\kern-.35em Z}}
\newcommand{\bP}{{\rm I\kern-.15em P}}
\newcommand{\Q}{\kern.3em\rule{.07em}{.65em}\kern-.3em{\rm Q}}
\newcommand{\R}{{\rm I\kern-.15em R}}
\newcommand{\h}{{\rm I\kern-.15em H}}
\newcommand{\C}{\kern.3em\rule{.07em}{.55em}\kern-.3em{\rm C}}
\newcommand{\T}{{\rm T\kern-.35em T}}

\begin{document}

\openup 1.5\jot

\centerline{INVESTIGATION OF THE INTERIOR OF COLORED BLACK HOLES AND THE EXTENDABILITY}
\centerline{OF SOLUTIONS OF THE EINSTEIN-YANG/MILLS EQUATIONS DEFINED IN THE FAR FIELD}

\vspace{.25in}
\centerline{by J.A. Smoller$^1$ and A.G. Wasserman}
\vspace{.50in}

\noindent
\underline{Abstract.}
We prove that any solution to the spherically symmetric $SU(2)$ 
Einstein-Yang/Mills equations that is defined in the far field and is 
asymptotically flat, is globally defined. This result applies in particular 
to the interior of colored black holes.

\bigskip

\section{Introduction.}
 \ \ \ \ \ In this paper we prove the following surprising property of 
 spherically symmetric solutions to the $SU(2)$ Einstein-Yang/Mills equations: 
  Any solution to the EYM equations which is defined in the far field $(r >> 
  1)$ and has finite (ADM) mass,
is defined for all $r > 0$.  We note that this is not true in the ``other
 direction"; i.e., if a solution is defined near $r=0$ with particle-like 
 boundary conditions, a singularity can develop at some $\rho > 0$, and the 
 solution cannot be extended for 
 $r > \rho$, (see [9, Thm. 4.1]).  Moreover, in general for nonlinear 
 equations, existence theorems are usually only local, with perhaps global 
 existence only for special parameter values .  However for these equations 
 we prove here a global existence result for all solutions defined in a 
 neighborhood of infinity.  Furthermore, we know (see [9]), that given any 
 event horizon $\rho > 0$, there are an infinite number of black-hole 
 solutions having event horizon $\rho$.  Our results in this paper imply 
 that all of these solutions can be continued back to $r=0$.  In particular,
  this gives information as to the behavior of the Einstein  metric and the 
  Yang-Mills field inside a black hole, a subject of recent interest; 
  see [4,5].

\bigskip
\noindent
\underline{\ \ \ \ \ \ \ \ \ \ \ \ \ \ \ \ \ \ \ }

$^1$ Research supported in part by the N.S.F., Contract No. DMS-G-9501128.

In the papers [10, 14], we have studied solutions defined in a neighborhood 
of $r=\infty$, and we proved that either the solution is defined up to 
some $r = \rho > 0$, in which case it is a black-hole solution of radius
 $\rho$, (as discussed in [9], and therefore continues through the event 
 horizon; i.e., to $\rho - \varepsilon \le r \le \rho$), or else the solution 
 is defined all the way to $r=0$, and is particle-like or is 
 Reissner-Nordstr\"{o}m-Like.  In this paper, we complete our
  investigtions by analyzing the behavior inside the black hole; i.e., on 
  the interval $0 < r < \rho$, see [4,5] for a discussion of the behavior 
  near $r=0$.

In order to describe our results, we recall that for the spherically symmetric EYM equations, the Einstein metric is of the form
\begin{equation}
ds^2 = -AC^2 dt^2 + A^{-1} dr^2 + r^2(d\theta^2 + \sin^2 \theta d\phi^2),
\end{equation}
and the $SU(2)$ Yang-Mills curvature 2-form is
\begin{equation}
F = w' \tau_1 dr \wedge d\theta + w'\tau_2 dr \wedge(\sin \theta d\phi) - (1 - w^2)\tau_3 d\theta \wedge (\sin \theta d \phi).
\end{equation}
Here $A,C$ and $w$ are functions of $r$, and $(\tau_1, \tau_2, \tau_3)$ form a basis for the Lie algebra $su(2)$.  Using (1.1) and (1.2), the spherically symmetric $SU(2)$ EYM equations are (cf [1 - 14]):
\begin{equation}
rA' + (1 + 2w'^2)A = 1 - {(1 - w^2)^2 \over r^2} \ ,
\end{equation}
\begin{equation}
r^2 Aw'' + \left[ r(1 - A) - {(1-w^2)^2 \over r} \right] w' + w(1 - w^2) = 0,
\end{equation}
and
\begin{equation}
{C' \over C} = {2w'^2 \over r}.
\end{equation}
Notice that (1.3) and (1.4) do not involve $C$ so that the major part of our effort is to study the coupled system (1.3), (1.4).

We define the ``mass function" $\mu(r)$ by
$$   \mu(r) = r(1 - A(r)).  $$
If
\begin{equation}
\lim_{r \rightarrow \infty} \ \mu(r) \equiv \bar{\mu} < \infty ,
\end{equation}
the solution is said to have finite (ADM) mass.  Our main result in this paper can be stated  as
\smallskip

\noindent
\underline{Theorem 1.1}.  {\it Any solution to the spherically symmetric $SU(2)$ EYM equations defined in the far field any having finite (ADM) mass, is defined for all $r>0$.}

\smallskip
Equivalently, (see Proposition 4.1), we can restate our result as

\smallskip
\noindent
\underline{Theorem 1.2}.  {\it Any solution to the spherically 
symmetric $SU(2)$ EYM equations defined in the far field and 
having $A(\bar{r}) > 0$ for some $\bar{r} > 1$, is defined for all $r > 0$.}

\smallskip

We now give an outline of the proof.  Assume that the solution is defined for
 all \\ $r > r_0 > 0$; we then prove that the solution can be continued 
 through $r_0$; i.e., on an interval of the form 
 $r_0 - \varepsilon < r < \infty$, for some $\ve > 0$.  In 
order to get a handle on the solution we first prove that $A(r)$ has at most
 a finite number of zeros on the interval $r_0 \le r < \infty$; this is the 
 main content of $\S3$.  Thus $A(r)$ must be of one sign for
   $r$ near $r_0, \ r > r_0$, and so there are
two cases to consider in the proof: $A > 0$ near $r_0$ or $A < 0$ near $r_0$.

When $A > 0$ near $r_0$, there are certain simplifying features of the problem; 
for example, $\mu'(r) > 0$ so $\mu(r)$ has a limit at $r_0$, and thus 
$\lim_{r\searrow r_0}A(r)$ exists.  If $A(r_0) \ge 1$ then $(A,w)$ is a 
Reissner-Nordstr\"{o}m-Like (RNL)
solution, and it was proved in [14] that such solutions are defined on 
$0 < r \le r_0$.  If $A(r_0) > 0, \ w^2(r_0) > 1$, and $(ww')(r_0) \ge 0$, 
this contradicts our assumption that the solution is defined in the far 
field, [10].  If $A(r_0) > 0$, $w^2(
r_0) > 1$, and $(ww')(r_0) < 0$, then it was proved in [14] that again 
the solution is an RNL solution.  Thus, in the case where $A > 0$ near $r_0$, 
we may assume that $1 > A(r) > 0$ and $w^2(r) < 1$ for $r$ near $r_0$.  
In this case, the results in [10] 
show that the solution can be continued beyond $r_0$; see Theorem 4.2.

The main thrust of this paper is to consider the case when $A(r) < 0$ for $r$ 
near $r_0$, $(r > r_0)$, and to prove that in this case too the solution can 
be continued beyond $r_0$. 
\vfill\eject

If $A < 0$ near $r_0$, there are two cases to consider: (I) Near $r_0$, $A$ 
is bounded  away from zero, and (II),  $A$ is not bounded away from zero; 
i.e., there is a sequence $r_n \searrow r_0$ such that $A(r_n) \rightarrow 0$. 
 In Case (I), we prove that the equations are non-singular at $r_0$, and thus 
 the solution can be continued beyond $r_0$.  In Case (II), the equations are 
 singular at $r_0$.  However, we prove in this case that these solutions are 
 exactly solutions of the type considered in [9], and the existence and 
 uniqueness theorems proved in [10] imply that the solution can be continued 
 beyond $r_0$.  These cases form the subject of $\S5$.

In $\S2$ we introduce some auxiliary functions which will be used in the paper, 
and we also recall some known results.  The reader is advised to consult this 
section as needed.  

The final section consists of a list of miscellaneous results, open questions and conjectures.   
\vfill\eject

\section{Preliminaries.}
\setcounter{equation}{0}

\ \ \ \ \ The static, spherically symmetric EYM equations, with gauge group $SU(2)$, can be written in the form (c.f. [1, 3, 7]):
\begin{equation}
rA' + (1 + 2w'^2) = 1 - {u^2 \over r^2},
\end{equation}
\begin{equation}
r^2Aw'' + \left[ r(1 - A) - {u^2 \over r} \right] w' + uw = 0,
\end{equation}
\begin{equation}
{C' \over C} = {2w'^2 \over r} ,
\end{equation}
where
\begin{equation}
u = 1 - w^2 .
\end{equation}
Here $w(r)$ is the connection coefficient which determines the Yang/Mills field, and $A$ and $C$ are the metric coefficients in (1.1).

If we define the function $\Phi$ by
\begin{equation}
\Phi(A, w, r) = r(1 - A) - {u^2 \over r} ,
\end{equation}
then (2.1) and (2.2) can be written in the compact form
\begin{equation}
rA' + 2Aw'^2 = \Phi/r,
\end{equation}
\begin{equation}
r^2Aw'' + \Phi w' + uw = 0.
\end{equation}
If $(A(r), w(r))$ is a given solution of (2.1), (2.2), then we write
\begin{equation}
\Phi(r) = \Phi \left( A(r), w(r), r \right).
\end{equation}
We note that (c.f. [8]) the function $\Phi$ satisfies the equation
\begin{equation}
\Phi ' = {2u^2 \over r^2} + 2Aw'^2 + {4uww' \over r}.
\end{equation}

We shall have occasion to analyse the behavior of the functions $v, f$ and $\mu$ defined by 
\begin{equation}
v = Aw' \ ,
\end{equation}
\begin{equation}
f = Aw'^2
\end{equation}
and
\begin{equation}
\mu = r(1 - A).
\end{equation}
These satisfy the respective equations ([8, 9])
\begin{equation}
v' + {2w'^2 \over r} + {uw \over r^2} = 0 ,
\end{equation}
\begin{equation}
r^2f' + (2rf + \Phi)w'^2 + 2uww' = 0,
\end{equation}
and
\begin{equation}
\mu ' = 2Aw'^2 + {u^2 \over r^2}.
\end{equation}

We now shall recall some results from the papers ([8-10, 12-14]); these will be needed in our development.

The first theorem gives us control on orbits which leave the region $w^2 < 1$.

\smallskip
\noindent
\underline{Theorem 2.1}. ([10, 14]): {\it Let $(A(r), w(r))$ be a solution of (2.1), (2.2), and assume that for some $r_0 > 0$, $w^2(r_0) > 1$ and $A(r_0) > 0$.}
\begin{itemize}
\item[i)] If $(ww')(r_0) > 0$, then there is an $r_1 > r_0$ such that $\lim_{r \nearrow  r_1}$ $A(r) = 0$, and $w'$ is unbounded near $r_1$.
\item[ii)]  If $ (ww') (r_0) < 0$, then there is an $r_1$, $0 < r_1 < r_0$ such that $A(r_1) = 1, \ A(r) > 0$ if $0 < r \le r_0$, and $\lim_{r \searrow 0} (A(r), w(r), w'(r)) = (\infty, \bar{w}, 0)$, for some $\bar{w}$. 
\end{itemize}
\smallskip

A solution which satisfies $A(r) > 1$  for some  $r > 0$, is called a \underline{Reissner-Nordstr\"{o}m-Like} (RNL) solution; see [14] for a discussion of these RNL solutions.

The next two theorems disallow degenerate behavior of the function $A(r)$.

\smallskip
\noindent
\underline{Theorem 2.2}. ([12,13]).  {\it Suppose that $(A(r), w(r))$ 
is a solution of (2.1), (2.2), and 
$\lim_{r \searrow \bar{r}} A(r)= 0$ $ = \lim_{r \searrow \bar{r}} A'(r)$. 
 Assume too that $A(r_1) > 0$ for some $r_1 > \max (\bar{r}, 1)$.  
 Then $(A,
w)$ is the extreme Reissner-Nordstr\"{o}m (ERN) solution: 
$A(r) = \left( {r-1 \over r} \right)^2, \\ w(r) \equiv 0$.}

\smallskip
\noindent
\underline{Theorem 2.3}. . {\it Suppose $(A(r), w(r))$ is any solution of 
(2.1), (2.2), defined on an interval $r_1 \le r \le r_2$, and set 
 $w_1 = w(r_1),  w_2 = w(r_2)$, and $M = \sup |Aw'^2(r)|$  for $r_1 \le 
 r \le r_2$.  Suppose $w_1 \le w(r) \le w_2$
for  $r_1 \le r \le r_2$, and suppose further that there is a 
 constant $\delta > 0$ such that $|\Phi(r)| \ge \delta$ on this $r$-interval.  
 Then there exists a constant $\eta > 0$, depending only 
 on $\delta, M,$ and $|w_1 - w_2|$ such that $|r_1 - r_2| \
ge \eta$.}

We next recall the notions of particle-like and black hole solutions of 
the EYM equations.

A (Bartnik-McKinnon) \underline{particle-like} solution of (2.1), (2.2) is 
a solution defined for all $r \ge 0, \ A(0) = 1$, $(w^2(0), w'(0)) = (1,0)$, 
and $w''(0) = -\lambda < 0$ is a free parameter; particle-like solutions are 
parametrized by (a discrete set of) $\lambda$: $(A(r,\lambda), w(r,\lambda))$.

\smallskip
\noindent
\underline{Theorem 2.4}. ([8,9,3]). {\it There is an increasing 
sequence $\la_n \nearrow \bar{\lambda} \le 2$, where 
$w(0, \la_n) = -\la_n$, such that the corresponding solutions 
$(A(r,\la_n), w(r,\la_n))$ are particle-like and 
 \\ $\lim_{r \rightarrow \infty} \left(A(r,\la_n), C(r_1, \la_n)\right), 
 \ w^2(r,\la_n), w'(r,\la_n)  = (1,1,1,0)$, and $\mu_n \equiv 
 \\ \lim_{r \rightarrow \infty} r\left( 1 - A(r,\la_n) \right) < \infty$.  
 Moreover,  $w(r,\la_n)$ has precisely n-zeros. }
\smallskip

A \underline{black-hole} solution of radius $\rho > 0$ of (2.1), (2.2) 
is a solution defined for all $r > \rho$, $\lim_{r \searrow \rho}A(r) = 0, 
A(r) > 0$ if $r > \rho$.  It was shown in [9] that the functions $A$ and $w$ 
are analytic at $\rho$, and that
 $(w(\rho), w'(\rho))$ lies on the curve $C_\rho$ in the $w-w'$ plane given by
$$	C_\rho = \{ (w,w') : \Phi(0, w, \rho)w' + uw = 0 \} . $$

The curves $C_\rho$ differ depending on whether $\rho < 1$, $\rho = 1$, 
or $\rho > 1$; these are depicted in Figures 1-3.

On each of these figures we have indicated the sign of $\Phi(\rho)$ in 
the relevant regions by $+$ or $-$ signs.  The components of $C_\rho$ for 
which $\Phi > 0$ correspond to (local) solutions for which $A'(\rho) > 0$, 
and (some) yield black-hole solutio
ns.  The other components correspond to (local) solutions with $A'(\rho) < 0$. 
 Black-hole solutions can only emanate from the component of the curve 
 containing $Q$ (c.f. Figures 1-3).  
 The obits through $P$ and $R$ have $A(r) < 0$ for some $r > \rho$. 
  Finally, we showed in [14] that the orbits through $R$ correspond to RNL 
  solutions.

Black hole solutions are parametrized by $w(\rho)$, and the relevant theorem 
for black solutions is:

\noindent
\underline{Theorem 2.5}. ([9]). {\it Given any $\rho > 0$, there is a sequence 
$(\alpha_n, \beta_n) \in C_\rho$, where $\Phi(\alpha_n, \rho, 0))  \not= 0$, 
such that the corresponding solution $(A(r,\alpha_n), w(r, 
\alpha_n), w'(r,\alpha_n)$ of (2.1),(2.2) is defined for all $r > \rho$ 
satisfies $A(r,\alpha_n) > 0$, and $w^2(r, \alpha_n) < 1$.  
Moreover, $\lim_{r \rightarrow \infty} ( A(r,\alpha_n), 
\\ w^2(r,\alpha_n),$ $w'(r,\alpha_n)) = (1,1,0)$,
 $\lim_{r \rightarrow \infty}  r( 1 - A(r, \alpha_n)) < \infty$ and $w(r, 
 \alpha_n)$ has precisely $n$-zeros.}

Our final result classifies solutions which are well-behaved in the far-field.  It does \underline{not} describe the behavior of either the gravitational field or the YM field, inside a black hole -- this is the subject dealt with in this paper.

\noindent
\underline{Theorem 2.6}. ([14]).   {\it Let $(A(r), w(r))$ be a solution of (2.1), (2.2) which is defined and smooth for $r > \bar{r} > 0$ and satisfies $A(r) > 0$ if $r > \bar{r}$.  Then every such solution must be in one of the following classes:}
\begin{itemize}
\item[(i)] $A(r) > 1$  for all  $r > 0$;
\item[(ii)] \underline{Schwarzschild Solution}:  $A(r) = 1 - {2m \over r}, \ w^2 \equiv 1, \ (m = \; {\rm const.})$;
\item[(iii)] \underline{Reissner-Nordstr\"{o}m Solution}: $A(r) = 1 - {c \over r} + {1 \over r^2}, \ w(r) \equiv 0, \ (c = \; {\rm const.})$;
\item[(iv)] \underline{Bartnik-McKinnon Particle-Like Solution};
\item[(v)] \underline{Black-Hole Solution};
\item[(vi)] \underline{RNL Solution}.
\end{itemize}

In each case, $\lim_{r\ra \infty} w^2(r) = 1$ or $0$ ($0$ only for RN solutions), $\lim_{r \ra \infty} rw'(r) = 0$ and $\lim_{r\ra \infty} A(r) = 1$.  The solution also has finite (ADM) mass; i.e. $\lim_{r\ra \infty} r(1 - A(r)) < \infty$.
\vfill\eject

\section{The zeros of A.}
\setcounter{equation}{0}

\ \ \ \ \ In this section we shall prove that the zeros of $A(r)$ are discrete, 
except possibly for an accumulation point at $r=0$.  We shall also show that 
$A$ can have at most two zeros in the region $r \ge 1$.  In proving these, 
we shall make use of Figures 1-3.

In the remainder of this paper we shall always assume that the following 
hypothesis (H) holds for a given solution $(A(r), w(r))$ of (1.3) and (1.4)

\begin{description}
\item[(H)] There is an $r_1 > 1$ such that the solution $(A(r), w(r))$ is 
defined for all $r > r_1$,  and $A(r_2) > 0$  for some $r_2 \ge r_1$.
\end{description}

\smallskip
\noindent
\underline{Theorem 3.1}.  {\it If the hypothesis $(H)$ holds, then $A$ has at 
most a finite number of zeros in any interval of the form $\ve \le r < \infty$, 
for any $\ve > 0$.  Furthermore, all the zeros of $A$, with at most two 
exceptions, lie in the set $r < 1$.}

\smallskip

Note that from [12], if $A(\bar{r}) = 0 = A'(\bar{r})$, for some $\bar{r} > 0$, then the solution is the extreme Reissner-Nordstr\"{o}m (ERN) solution
$$    A(r) = \left( {r-1 \over r} \right)^2 , \ \ \ \ w(r) \equiv 0.  $$
For this solution, Theorem 3.1 clearly is valid.  Thus, in this section we 
shall assume that if $A(\bar{r}) = 0$, then $A'(\bar{r}) \not= 0$.  
In this case from$^2$ ([10]), $\lim_{r \searrow \bar{r}} A(r) = 0$,  
$\lim_{r \searrow \bar{r}} (w(r), w'(r))$ $
= (\bar{w}, \bar{w}')$ exists, and $(\bar{w}, \bar{w}') \in C_{\bar{r}}$; 
(c.f. Figures 1-3).  

\smallskip
\noindent
{Proposition 3.2}. {\it A cannot have more than two zeros in the region $r \ge 1$.}

\noindent
\underline{\ \ \ \ \ \ \ \ \ \ \ \ \ \ \ \ \ \ \ \ }

$^2$ In ([10]), the result was demonstrated for the case where $A(r) > 0$ for $r$ near $\bar{r}$, $r > \bar{r}$, but the same proof holds if $A(r) < 0$ for $r$ near $\bar{r}, r > \bar{r}.$
\vfill\eject

\noindent
\underline{Proof.}  Suppose that $A$ has 3 zeros in the region $r \ge 1$. 
 Then there must exist $\rho, \eta, \ 1 \le \eta < \rho$ with $A(\rho) = 
 0 = A(\eta)$ and $A'(\eta) < 0 < A'(\rho)$; c.f. Figure 4.  
 Since $(w(\eta), w'(\eta)) \in C_\eta$ and $\eta
> 1$, we see from Figure 3 that $w^2(\eta) > 1$.  Then from Theorem 2.1,(ii),
 $A$ cannot have any zeros if $r < \eta$.  This contradiction establishes the 
 result.   []We next prove 

\noindent
\underline{Proposition 3.3}. {\it If $0 < r_0 <1$, then $r_0$ cannot be a 
limit point of the zeros of $A$.}

\smallskip
Notice that Theorem 3.1 follows at once from Propositions 3.2 and 3.3.

\noindent   
\underline{Proof.}  We shall show that there is a neighborhood of $r_0$ 
in which $A \not= 0$.  

Choose $\ve > 0$ such that $r_0 + \ve < 1$.  We will show, using Theorem 2.3, 
that there exists an $\eta > 0$ such that if $z_1$ and $z_2$ are two 
consecutive zeros of $A$, $\ r_0 < z_1 < z_2 < r_0 + \ve < 1$,
\begin{equation}
A(z_1) = 0 = A(z_2), \ \ \ \ A'(z_1) > 0 > A'(z_2)
\end{equation}
then
\begin{equation}
z_2 - z_1 > \eta.
\end{equation}
This implies that there can be at most a finite number of zeros of $A$ in 
the interval $(r_0, r_0+\ve)$.

Now $A(z_2) = 0$ implies that $(w(z_2), w'(z_2))$ lies on $C_{z_2}$, and 
$A'(z_2) < 0$ implies that $(w(z_2), w'(z_2))$ lies on the middle curve in 
Figure 1, (where $\rho$ is replaced by $z_2$).  Without loss of generality, 
assume $w'(z_2) > 0, \ w(x_2) >0$.

Now define $\delta$ by
$$ (r_0 + \ve) - {1 \over (r_0 + \ve)} = -2\delta < 0.   $$
Then there exists a constant $c > 0$ such that
$$ (r_0 + \ve) - {u^2 \over (r_0 + \ve)} \le -\delta < 0, \ \ \ 
{\rm if} \ \ \ |w| < c.   $$
Hence
$$    r - {u^2 \over r} \le -\delta , \ \ \ {\rm if} \ \ \ r_0 
\le r \le r_0 + \ve ,   $$
and thus
\begin{equation}
\Phi(r) = r - {u^2 \over r} - rA < -\delta \ , \ \ \ \ z_1 \le r \le z_2 \ ,
\end{equation}
since $A(r) > 0$ if $z_1 < r < z_2$.  Let $w_1 = -c, \ w_2 = 0$; then there 
exist $r_1,r_2, \ \  z_1 < r_1 < r_2 < z_2$ such that $w(r_1) = -c$, $w(r_2)
 = 0$.  (This is because $A$ cannot change sign in the interval $-c \le w \le 
 0$; c.f. Figure 1, and Figure 5.)

Now in view of (3.3), if we can show that there is an $M > 0$, $M$ 
independent of $z_1,z_2$ for which
\begin{equation}
|Aw'^2| \le M, \ \ r_1 \le r \le r_2 \ \ \ ({\rm equivalently}, 
\ \ w_1 \le w(r) \le w_2),
\end{equation}
then on the interval $w_1 \le w \le w_2$, we may apply Theorem 2.3 to
 conclude that (3.2) holds.  Thus the proof of Proposition 3.3 will be 
 complete once we prove (3.4); this is the content of the following lemma.

\smallskip
\noindent
\underline{Lemma 3.4}.  {\it If $-c \le w(r) \le 0$, then}
\begin{equation}
f(r) \equiv (Aw'^2)(r) < {2 \over r^2_0}.
\end{equation}
\noindent
\underline{Proof of Lemma 3.4}. Recall that $f$ satisfies
\begin{equation}
r^2f' + (2rf + \Phi)w'^2 + 2uww' = 0.
\end{equation}
We first claim that there is a value $a < r_1$, with $f(a) = 0$, and for $a 
\le r < r_1$, $-1 \le w(r) \le -c$, and $w'(r) \ge 0$.  Indeed, note that the 
orbit cannot exit the region $w^2 < 1$ through $w=-1$, for $r > z_1$, because
 by Theorem 2.1 (ii), there would be no zero of $A$ smaller than $r_1$.  
 Therefore, either the point $\left( w(z_1), w'(z_1)\right)$ lies in $-1 
 \le w \le -c$, $w' > 0$, in which case we take $a = z_1$, or else the orbit 
 crosses the segment $-1 \le w \le -c$, $w' = 0$ at some $r=a$, and again 
 $f(a) = 0$.

We now prove 
\begin{equation}
{\rm if} \ \ f(r) = {2 \over r^2_0}\, , \ \ \ \ \ {\rm then}  \ \ \ \ f'(r) < 0.
\end{equation}
for $r$ in the interval $(a,r_2)$.  Since $f(a) = 0$, then if (3.7) holds, there can be no first value of $r$ for which $f(r) = {2 \over r^2_0}$, and hence (3.5) holds.  Thus it suffices to prove (3.7).

To do this, we first note that
\begin{equation}
\Phi(r) \ge - {1 \over r_0} \ , \quad {\rm if} \quad a < r < r_2.
\end{equation}
Indeed
$$ \Phi(r) = r(1-A) - {u^2 \over r} \ge - {u^2 \over r} \ge - {1 \over r} \ge - {1 \over r_0}. $$
Now from (2.14), we have, when $f = {2 \over r^2_0}$,
\begin{eqnarray}
r^2f'(r) &=& \left[ -(2rf + \Phi)w' - 2uw \right]w' \nonumber \\
&=& \left[ -\left(2r {2 \over r^2_0} + \Phi\right)w' - 2uw\right]w' \nonumber \\
&\le& \left[ \left( - {4r \over r_0} {1 \over r_0} + {1 \over r_0} \right) w' + 2 \right]w' \\
&=& \left[ 2 + {w' \over r_0} \left(1 - {4r \over r_0} \right) \right]w' \nonumber \\
&\le& \left[ 2 - {3w' \over r_0} \right]w' \nonumber ,
\end{eqnarray}
where we have used (3.8).  Now when $f = {2 \over r^2_0}\,$, $w'^2 = {2 \over r^2_0 A} > {2 \over r^2_0}$\,, or $w' > {{\sqrt 2} \over r_0}$.  Using this in (3.9) gives
$$   r^2f' \le \left( 2 - {{3\sqrt{2}} \over r^2_0} \right)w' < (2 - 3\sqrt{2} )w' < 0,    $$
and this gives (3.7).  Thus the proof of Lemma 3.4 is complete, and as we have seen, this proves Proposition 3.3.\ \ \   []
\vfill\eject

\section{The case   $A > 0$ near $r_0$.}
\setcounter{equation}{0}

\ \ \ \ \ In this section we shall first prove the equivalence of 
Theorems 1.1 and 1.2.  Then we shall prove Theorem 1.2 in the case 
where $A(r) > 0$ for $r$ near $r_0, \; r > r_0$.  In view of Theorem 3.1, 
we know that $A$ can have at most a finite number of zeros 
on the interval $(r_0, \infty)$.  Hence $A(r)$ is of one sign for 
$r > r_0$  In this section we shall prove that if $A(r) > 0$ for $r$ 
near $r_0$, the solution can be extended.  The far more difficult case where 
$A(r) < 0$ for $r$ near $r_0$, w

ill be considered in $\S 5$.

\smallskip
\noindent
\underline{Proposition 4.1.} {\it Theorems 1.1 and 1.2 are equivalent.}

\noindent
\underline{Proof.}  Assume that Theorem 1.1 holds, and that $A(\bar{r}) > 0$, 
where $\bar{r}$ is as given in the statement of Theorem 1.2.  
Consider $\left( w(\bar{r}), w'(\bar{r}) \right)$.  If $w^2(\bar{r}) 
\ge 1$ and $(ww')(\bar{r}) > 0$, then from Theorem 2.1, i), the solution 
cannot exist for all $r > \bar{r}$, and this contradicts our assumptions.  
If $w^2(\bar{r}) \ge 1$ and $(ww')(\bar{r}) < 0$, then from Theorem 2.1,  
ii), the solution is an RNL solution and is thus defined for all 
$r, \ 0 < r < \bar{r}$.  Thus, we may assume that $w^2(\bar{r}) < 1$.  
If $w^2(\tilde{r}) > 1$ for some $\tilde{r} > \bar{r}$, then $(ww')
(\tilde{r}) > 0$, so again Theorem 2.1,  i) implies that the solution is 
not defined in the far-field.  Hence we may assume that the 
orbit stays in the region $w^2(r) < 1$ for all $r > \bar{r}$.  
Moreover $A(r) > 0$ for all $r > \bar{r}$ because $A(r) = 0$ for some 
$r > \bar{r} > 1$ cannot occur.  (In $w^2 < 1$, ``crash" can occur only 
if $r < 1$; see [7].)  Thus from [14, Propos. 6.2], 
$\lim_{r \ra \infty} \mu(r) < \infty$, hence Theorem 1.2 holds.

Conversely, if Theorem 1.2 holds, then (1.6) implies that $\lim_{r\ra \infty} 
r(1 - A(r)) < \infty $ so $A(r) \ra 1$ as $r \ra \infty$; in particular $A(r)
 > 0$ for $r$ large.  This implies that Theorem 1.1 holds. \ \ \ \ []

This last result, this justifies our assumption that in the remainder of 
this paper that the following hypothesis (H) holds for a given solution 
$(A(r), w(r)$ of (1.3) and (1.4):

\begin{description}
\item[(H)] There is an $r_1 > 1$ such that the solution $(A(r), w(r))$ is defined for all $r > r_1$,  and $A(r_2) > 0$  for some $r_2 \ge r_1$.
\end{description}

\vfill\eject

We now let $r_0$ be any given positive number, and assume that the solution $(A(r), w(r))$, of (1.3), (1.4) is defined for all $r > r_0$.  We then have the following theorem.

\smallskip
\noindent
\underline{Theorem 4.2}.  Assume that hypothesis (H) holds, and that $A(r) > 0$ for $r$ near $r_0\;, \ r > r_0$.  Then the solution can be extended to an interval of the form $r_0 - \ve < r \le r_0$.

\noindent
\underline{Proof.}  It follows from Proposition 2.6 that either $w^2(r) 
< 1$ for all $r$ near $r_0$\,, or else $(A,w)$ is an RNL solution and is 
thus defined for $0 < r \le r_0$.  In the case $w^2(r) < 1$ for all $r$ 
near $r_0\,$ then if $A(r)$ is bounded
 away from zero for $r$ near $r_0\,$ the solution must continue into 
 a region of the form $(r_0-\ve, r_0]$, for some $\ve > 0$. (The proof 
 of this fact is the same if $A > 0$ or $A < 0$  near $r_0$.  In (5.6) 
 below we give the proof for $A < 0$, so we omit the proof here). 
  If, on the other hand, $A$ is not bounded away from zero near $r_0$, 
  then $A(r_n) \rightarrow 0$ for some sequence $r_n \searrow r_0$.  
  In [10], we have shown that this implies $\lim_{r \searrow r_0} A(r) = 0$, 
  and $\lim_{r \searrow r_0} (w(r), w'(r)) \in C_{r_0}$ so the 
  solution $(A,w)$ is analytic at $r_0$ and thus again continues 
  past $r_0$; i.e., to an interval of the form $r_0 - \ve \le r \le r_0$.  
  This completes the proof of Theorem 4.2. \ \ \ \ \ []

\bigskip

In the next section we shall consider the case where $A(r) < 0$ for $r$ near
 $r_0, \ r > r_0$,

\setcounter{section}{4}
\section{The case $A<0$ near $r_0$.}

\ \ \ \ In this section we assume that the solution $(A,w)$ of (1.3), (1.4) is 
defined for all $r > r_0$, and that $A(r) < 0$ for $r$ near $r_0\,, \ r > r_0$. 
 We shall prove that the solution can be continued past $r_0$ .  This is the 
 content of the following theorem.

\smallskip
\noindent
\underline{Theorem 5.1}. {\it Assume that hypothesis (H) holds and that $A(r) < 0$ for $r$ near $r_0\,, \ r > r_0$.  Then the solution can be continued to an interval of the form $r_0 - \ve < r \le r_0$.}

\smallskip

Notice that Theorems 4.1 and 5.1 imply Theorem 1.2.

\noindent
\underline{Proof.}  There are two cases to consider:

\underline{Case 1.}  There are positive numbers $\delta$ and $\Delta$ such that
\begin{equation}
A(r) < -\delta \ , \ \ \ \ {\rm if} \ \ \ \ 0 < r_0 < r < r_0 + \Delta \ ;
\end{equation}

\underline{Case 2.}  There is a $\Delta > 0$ such that
\begin{equation}
A(r) < 0 \ , \ \ \ \ {\rm if} \ \ \ \ 0 < r_0 < r < r_0 + \Delta \ ;
\end{equation}
and for some sequence $r_n \searrow r_0$,
\begin{equation}
A(r_n) \rightarrow 0.
\end{equation}

We begin the proof of Theorem 5.1 by first considering Case 1.  We shall need a few preliminary results, the first of which is

\noindent
\underline{Lemma 5.2}.  {\it If (5.1) holds, and $w(r)$ is bounded near $r_0\ \ (r > r_0)$, then $w'(r)$ is bounded near $r_0$.}

\noindent
\underline{Proof.}  From (2.7), we can write
\begin{equation}
w'' + {\Phi \over r^2A} \; w' = - {uw \over r^2A}.
\end{equation}
Since
$$  {\Phi \over r^2A} = {1 \over rA} - {1 \over r} - {u^2 \over r^3A} $$
we see that both $\Phi/r^2A$ and $uw/r^2A$ are bounded near $r_0$.  Thus the coefficients in (5.1) as well as the rhs are bounded, so $w'$ too is bounded near $r_0$. 
\ \ \ \ []

\medskip
\noindent
\underline{Lemma 5.3}.  {\it If $w'$ is bounded near $r_0$, then $A$ is bounded near $r_0$.}

\noindent
\underline{Proof.}  From (2.1), we have
\begin{equation}
rA' + (1 + 2w'^2)A = 1 - {u^2 \over r^2}.
\end{equation}
The hypothesis implies that $w$ is bounded near $r_0$ so the coefficients 
of (5.5) are bounded near $r_0$.  Thus $A$ too is 
bounded near $r_0$. \ \ \ \ []

\medskip

These last two results enable us to dispose of the case where (5.1) holds, and also 
\begin{equation}
w(r) \ \ {\rm is \ bounded \ near} \ \ r_0(r > r_0).
\end{equation}
Since (5.6) holds, then $A,w$, and $w'$ are bounded near  $r_0$, and by 
(5.1), $A(r) < - \delta$, we see from (1.3), and (1.4), that $A', w'$ and 
$w''$ are bounded.  Thus $\lim_{r\searrow r_0}(A(r), w(r), w'(r), r)$ 
$(\bar{A}, \bar{w}, \bar{w}', r_0) 
\equiv P$ exists where $\bar{A} > 0$.  Hence the orbit through $P$ is thus defined
 on an interval , $r_0 - \ve < r < r_0 + \ve$, for some $\ve > 0$.

\smallskip
\noindent
\underline{Remark.}  We did not use the fact that $A < 0$ to obtain this
 conclusion; all we needed was $A$ bounded away from $0$ and $w$ bounded
  near $r_0$.

We shall now show that in Case 1, $w$ \underline{must} be bounded near $r_0$. 
 To do this, we will assume that $w$ is unbounded near $r_0,\; \ r > r_0$, 
 and we shall arrive at a contradiction.

Thus, assume that for some $\ve > 0$,
\begin{equation}
w(r) \ {\rm is \ unbounded \ on} \ \ (r_0\;, \ r_0 + \ve).
\end{equation}

\vfill\eject
\noindent
\underline{Lemma 5.4}. {\it If $A(r) < 0$ for $r$ near $r_0$, and (5.7) holds,
 then the projection of the orbit $(w(r), w'(r))$ has finite rotation about 
 $(0,0)$, and about $(\pm 1,0)$ for $r$ near $r_0$.}

\smallskip
\noindent
\underline{Remark.}  Note that we do not assume (5.1) but only that $A < 0$ 
near $r_0$.  In Case 2, we use the contrapositive of Lemma 5.4; i.e., if $A < 
0$ for $r$ near $r_0$, and if the orbit has infinite rotation about either 
$(0,0)$ or $(\pm 1,0)$, then $w$ is bounded near $r_0$.

\noindent
\underline{Proof.}  Assume that the orbit has infinite rotation about either 
$(0,0)$, or $(\pm 1, 0)$; we will show that this leads to a contradiction.

Since (5.7) holds, the orbit must rotate infinitely many times outside the region $w^2 \le 1$, as $r \searrow r_0$.  We may also assume without loss of generality that 
$\underline{\lim}_{r\searrow r_0} w(r) = - \infty$.  It follows that there 
exists sequences $\{r_n\}, \{s_n\}$, $r_{r+1} < s_{n+1} < r_n$, with 
$w'(r_n) = 0$, $w(s_n) = -2$, $\lim w(r_n) = -\infty$, and $w(r_n)
 < w(r) < w(s_n)$, for $r_n < r < s_n$;  c.f. Figure 6.
 
We first show that for $w(r) \le -2$, $w'$ is bounded; i.e, (as in the 
proof of Lemma 3.4, (c.f. (3.7)),
\begin{equation}
{\rm if} \ \ w'(r) = {2 \over 3} (r_0 + \ve), \ \ {\rm then} \ \ w''(r) < 0.
\end{equation}
To prove (5.8), we use (2.7):
\begin{equation}
w'' = {-uw \over r^2A} - {(r-rA)w' \over -r^2A} + {u^2w' \over r^3A} < {u \over r^3 A} 
[-rw + uw']. 
\end{equation}
Thus, if for some $r > r_0$, and $w(r) \le -2$, we had $w'(r) = {2 \over 3}(r_0 + \ve)$, then since ${w \over u} \le {2 \over 3}$, it follows that
$$    w'(r) = {2 \over 3} (r_0 + \ve) > {w \over u} r \ , $$
so that (5.9) implies (5.8).  Thus, if $w(r) \le -2$, then $w'(r) < {2 \over 3} (r_0 + \ve)$.
Since $s_n - r_n < \ve$, we have, for large $n$
$$   -2 \ - w(r_{n+1}) < {2\ve \over 3} (r_0 + \ve);$$
this violates (5.7).  \ \ \ \ []

\smallskip
\noindent
\underline{Corollary 5.5}.  {\it If (5.1) and (5.7) hold, then $\lim_{r\searrow r_0} |w(r)| =  \infty$.}

\noindent
\underline{Proof:}  For $r$ near $r_0$, the lemma implies that
 the orbit has finite rotation near $r_0$.  Thus the orbit must lie 
 in one of the four strips, $w < -1$, $-1 < w < 0$, $0 < w < 1$, $w > 1$.  
 Since in each strip $w''$ is of fixed sign when $w'
= 0$ it follows  then that $w'$ is of one sign near $r_0$, so that $w$ has a 
limit at $r_0$; since $w(r)$ is not bounded near $r_0$, the result follows. \ \ \ \ []

\medskip

It follows from the last result that if $w$ is unbounded near $r_0$, 
then the orbit must lie in either region (1) or region (5), as depicted 
in Figure 7.  We will assume that the orbit
lies in region (5) for $r$ near $r_0$; the proof for region (1) is similar, 
and will be omitted.  Thus, assuming (5.1), and (5.7) we have $w'(r) < 0$ 
near $r_0$, and 
\begin{equation}
\lim_{r\searrow r_0} w(r) = + \infty .
\end{equation}
Since $r_0$ is finite, (5.10) implies

\begin{equation}
w'(r) \ \ {\rm is\; unbounded\; for} \ \  r \ {\rm near} \  r_0 \ \ \ \ (r>r_0).
\end{equation}

\noindent
\underline{Lemma 5.6}.  {\it If $A(r) < 0$ for $r$ near $r_0 \ (r > r_0)$, and (5.10) holds, then}
\begin{equation}
\lim_{r \searrow r_0} w'(r) = - \infty.
\end{equation}
\smallskip
\noindent
\underline{Remark.}  We do not use hypothesis (5.1) in this lemma, but we only assume $A < 0$ near $r_0$.  This result will be used in Case 2.

\noindent
\underline{Proof.}  If $w'$ does not have a limit at $r_0$, then in view of (5.11), we can find sequences $r_n \searrow r_0, \ s_n \searrow r_0, \ \ r_n < s_n < r_{n+1}$, such that
\begin{equation}
w'(s_n) = -n, \ \ \ \ w'(r_n) = - {n \over 2},
\end{equation}
and
\begin{equation}
{\rm if} \  r_n \le r \le s_n, \  w'(r) \le - {n \over 2} .
\end{equation}
Then if $r_n \le r \le s_n$ and $n$ is large, (2.2) gives
\begin{eqnarray*}
-w''(r) &=& {(r-rA - {u^2 \over r})w' + uw \over r^2A} \\
          &=& {(-r^2A - {u^2 \over 2})w' + (ruw + r^2w' - {u^2 \over 2}\;w') \over r^3A} \\
          &<& {-w'(-r^2A - {u^2 \over 2}) \over -r^3A} \\
          &<& -w' \left( {-r^2A \over -r^3A} \right) = {-w' \over r} < {-w' \over r_0}. 
\end{eqnarray*}
Thus
\begin{equation}
{-w'' \over -w'} < {1 \over r_0} \ ,
\end{equation}
and so integrating from $r_n$ to $s_n$, gives
$$    \ell n2 = \ell n \left( {-w'(s_n) \over -w(r_n)} \right) < {1 \over r_0} (s_n - r_n),  $$
so that
\begin{equation}
s_n - r_n > r_0 \ell n 2.
\end{equation}
But for large $n$, $r_n < 1 + r_0$, so that (5.16) implies
$$    1 = (1 + r_0) - r_0 \ge \Sigma (s_n - r_n) = \infty.   $$
This contradiction establishes (5.12) and the proof of the lemma is complete. \ \ \ \ []

Thus to dispense with Case 1, and obtain the desired contradiction (assuming 
that $w$ is unbounded near $r_0$), we shall prove the following proposition.

\smallskip
\noindent
\underline{Proposition 5.7}. {\it It is impossible for (5.1) and (5.7) to hold.}

To prove this proposition, we shall obtain an estimate of the form
\begin{equation}
w''(r) \le k(-w'(r))
\end{equation}
for $r$ near $r_0$.  Integrating from $r > r_0$ to $r_1 > r$, gives
$$   \ell n\left( {-w'(r) \over -w'(r_1)} \right) \le k(r_1 - r)    $$
and this shows that $w'$ is bounded near $r_0$, thereby violating (5.12).

In order to prove (5.17), we need two lemmas, the first of which is

\smallskip
\noindent
\underline{Lemma 5.8}.  {\it If $A(r) < 0$ for $r$ near $r_0$, $(r > r_0)$, and both (5.10) and (5.12) hold, then writing $ Aw'^2 = f$, we have}
\begin{equation}
-f(r) > w(r)^5,  \ \  {\rm if} \ \ r \ \ {\rm is \ near} \ \  r_0.
\end{equation}

\smallskip
\noindent
\underline{Remark.}  We do not assume that (5.1) holds, but only that $A < 0$ near $r_0$.  This result too will be used in Case 2.

\noindent
\underline{Proof of Lemma 5.8.}  We write (2.14) in the form (c.f. (2.5))
\begin{equation}
r^2f' + rfw'^2 + (rf + r - rA)w'^2 +\left( {-u^2 \over r} w'^2 + 2uww' \right) = 0.
\end{equation}
Now for $r$ near $r_0$,
\begin{equation}
rf + r - rA = rAw'^2 + r - r A = rA(w'^2 - 1) + r \le 0,
\end{equation}
in view of (5.12).  Furthermore, if $r$ is near $r_0$,
\begin{equation}
{-u^2 \over r} w'^2 + 2uww' < 0
\end{equation}
because of (5.10), and (5.12).  Thus (5.19)-(5.21) imply $r^2f' + rfw'^2 > 0$, so that for $r$ near $r_0$
$$	f' > -f \left( {-w' \over r}\right) (-w') > fw',	  $$
or $f'/f < w'$.  Integrating from $r$ to $r_1$, where $r_0 < r < r_1$, and $r_1$ is close to $r_0$, gives
$$	\ell n(-f) \bigg|^{r_1}_r < w(r_1) - w(r) ,  $$
so that
$$  \ell n (-f(r)) > w(r) - k_1 \ ,   $$
where $k_1 = w(r_1) - \ell n (-f(r_1))$.  Exponentiating gives
$$	-f(r) > k_2 e^{w(r)} > w(r)^5 ,	$$
for $r$ near $r_0$, in view of Corollary 5.5. \ \ \ \ \ []

We shall use this last lemma for proving the following result.

\noindent
\underline{Lemma 5.9}.  {\it Assume that (5.1) and (5.10) hold.  Then there is a constant $k > 0$ such that}
\begin{equation}
-A(r) > kw(r)^4, \quad {\rm for}\   r \  {\rm near} \  r_0.
\end{equation}

\noindent
\underline{Proof.}  From (2.1), if $r$ is near $r_0$, 
$$
A' = {\Phi \over r^2} - {2 \over r} f \ge \left( {-u^2 \over r^3} - {f \over r} \right) - {f \over r} 
> {-f \over r},  $$
where we have used (5.18).  Thus, using (5.12),
$$    A' > {-Aw'^2 \over r} = k_3 Aw' , $$
for some $k_3 > 0$.  It follows that for some constant $k > 0$, 
$$	-A(r) > e^{k_3w} > kw^4 \ ,    $$
if $r$ is near $r_0$. \ \ \ \ []

\vfill\eject
We can now complete the proof of Proposition 5.7.  As we have seen earlier, it suffices to prove (5.17).  Now since we are in region (5) (c.f. Figure 7), $uw < 0$, so that for $r$ near $r_0$, (2.2) gives
$$
w'' < {(r - rA)w' - {u^2 \over r} w' \over -r^2A} \ < \ {u^2w' \over r^3A} < {u^2w' \over r^3_0A}
< c_1 {w^4w' \over A}
$$
where $c_1$ is a positive constant.  Thus, 
$$    w'' < {c_1w^4(-w') \over -A} \ \ < \ \ {c_1 \over k} (-w') \equiv c(-w')    $$
where we have used (5.22).  This proves (5.17), and as we have seen, completes the proof of Proposition 5.7. \ \ \ \ []

We now consider Case 2, where (5.2) and (5.3) hold; we will show that:
\begin{equation}
\lim_{r\searrow r_0} A(r) = 0 ,
\end{equation}
and
\begin{equation}
\lim_{r\searrow r_0} (w(r), w'(r) ) = (\bar{w}, \bar{w}') \in C_{r_0}.  
\end{equation}

\noindent
\underline{Remark.}  If (5.23) and (5.24) hold, then by the uniqueness theorem of [10], the solution is analytic at $r_0$ and hence continues past $r_0$. 

We begin with the following result.

\noindent
\underline{Proposition 5.10.}  {\it If (5.2) and (5.3) hold, then the orbit has finite rotation $\Omega$ about (0,0) in the $(w-w')$-plane; i.e. $\Omega < \infty$.}

The proof will follow from a series of lemmas, the first of which is

\noindent
\underline{Lemma 5.11}.  {\it Assume that (5.2) and (5.3) hold.  If the rotation $\Omega = \infty$, or if $w$ is bounded near $r_0$, then}
\begin{equation} 
\lim_{r \searrow r_0} \ A(r) = 0.
\end{equation}

\noindent
\underline{Proof.} If $\Omega = \infty$, then $w$ is bounded near $r_0$, by Lemma 5.4 and the remark following.  Thus we will prove that if $w$ is bounded near $r_0$, then (5.25) holds, or equivalently, that
\begin{equation}
\lim_{r \searrow r_0} \mu(r) \equiv \lim_{r \searrow r_0} r(1 - A(r)) = r_0 \ ,
\end{equation}
Since $A(r) < 0$ for  $r$ near $r_0$,
$$    \mu(r) = r(1 - A(r)) \ge r > r_0, \ {\rm if} \ r > r_0\ ,$$
so since $A(r_n) \ra 0$, 
\begin{equation}
\lim_{\overline{r \searrow r_0}} \mu(r) =r_0 .
\end{equation}
We shall next prove
\begin{equation}
\overline{\lim}_{r\searrow r_0} \  \mu(r) \le r_0 .
\end{equation}
and this together with (5.27) will prove (5.26).

If $\overline{\lim}_{r\searrow r_0} \  \mu(r) > r_0 $, then we can find numbers $b$ and $c$, $b > c > r_0$, and sequences $\{ s_n\}, \{t_n\}$, $r_0 < t_{n+1} < s_n < t_n$, with $\mu(s_n) = c$, $\mu(t_n) = b$.  Thus
$$ b-c = \mu(t_n) - \mu(s_n) = \mu '(\xi)(t_n - s_n),  $$
where $\xi$ is an intermediate point.  Now from (2.15) for $r$ near $r_0$,
$$ \mu '(r) = 2Aw'^2 + {u^2 \over r^2} \le {u^2 \over r^2} \le k ,$$
since $w$ is assumed to be bounded.  Hence $(b-c) < k(t_n - s_n)$, or $t_n - s_n > (b-c)/k > 0$.  This is a contradiction since $\sum_n (t_n - s_n)$  is finite.  Thus (5.28) holds and the proof is complete. \ \ \ \ []

\medskip

Combining Lemmas 5.4 and 5.11, we get as an immediate corollary,

\noindent
\underline{Corollary 5.12}.  {\it  If (5.2) and (5.3) hold, and $\Omega = \infty$, then $\Phi(r)$ is bounded for $r$ near $r_0$.}

We next have

\noindent
\underline{Lemma 5.13}.  {\it If (5.2) and (5.3) hold, and $w$ is bounded near $r_0$,  then either $Aw'^2$ is bounded near $r_0$, or $\lim_{r\searrow r_0} (Aw'^2)(r) = -\infty$.}

\noindent
\underline{Proof.}  We write $f = Aw'^2$, and again use (2.14):
\begin{equation}
r^2f' + (2rf + \Phi)w'^2 + 2uww' = 0.
\end{equation}
If $f$ is not bounded near $r_0$, then (Lemma 5.11) since $\Phi$ and $w$ are bounded, (5.29) shows that $f' > 0$ if $f$ is sufficiently large, and the result follows.   []

\noindent
\underline{Lemma 5.14}. {\it If (5.2) holds, and $Aw'^2$ is bounded near $r_0$, then the rotation number $\Omega$ is finite.}

\noindent
\underline{Proof.}  We are going to apply Theorem 2.3 with $w_1 = -1, 
\ w_2 = -1 + \ve$, for some $\ve > 0$.  Thus assume $\Omega = \infty$; then 
there exists a sequence $r^n_0 \searrow r_0$ with $w(r^n_0) = 0$,
 $w'(r^n_0) > 0$.  Since $A < 0$ near $r_0$,
the orbit cannot cross the segment $w' = 0$, $-1 \le w \le 0$ for $r < r^n_0$. 
 Thus we can find $\ve > 0$ and numbers $r^n_{-1}$, and $r^n_{-1 + \ve}$, 
 such that $w(r^n_{-1}) = -1$, $w(r^n_{-1 + \ve}) = -1 + \ve$, and 
 for $r^n_{-1} \le r \le r^n_{-1 + \ve}$, we have $-1 < w(r) < -1 + \ve$, 
 and for $r^n_{-1 + \ve} \le r \le r^n_0$, $-1 + \ve < w(r) < 0$.  
 By hypothesis, $Aw'^2$ is bounded near $r_0$, so in particular 
 on $r^n_{-1} \le r \le r^n_{-1 + \ve}$, for large $n$.  In order to apply 
 Theorem 2.3, it only remains to show that $\Phi(r)$ is bounded away 
 from $0$ on this interval if $\ve$ is small.

Choose $\ve > 0$ so small that
\begin{equation}
(1 - w^2)^2 < {1 \over 10} \ r^2_0, \ \ {\rm if} \ \ -1 \le w \le -1 + \ve.
\end{equation}
On this interval,
\begin{equation}
\Phi = r - rA - {u^2 \over r} > r - {u^2 \over r} > r_0 - {.1r^2_0 \over r_0} = .9r_0.
\end{equation}

Now by Theorem 2.3,  there exists an $\eta > 0$, such that for each $n$,
$$    r^n_{-1+\ve} - r^n_{-1} \ge \eta.    $$
But as $r^n_{-1+\ve}$ and $r^n_{-1}$ both lie in $(r_0, r_0 + 1)$ for large $n$, we have
$$   1 = (r_0 + 1) -r_0 \ge \Sigma \left( r^n_{-1+\ve} - r^n_{-1} \right) = \infty,  $$
and this is a contradiction.    []

Our final lemma in the proof of Proposition 5.10 is the following

\noindent
\underline{Lemma 5.15}. {\it If (5.2) and (5.3) hold, and $\Omega = \infty$,  $Aw'^2$ is bounded near $r_0$. }

\noindent
\underline{Proof.}   By Corollary 5.12, $\Phi$ is bounded.  From (2.6), if $Aw'^2 \rightarrow -\infty$, then as $r \searrow r_0$, 
$$     rA' = -2Aw'^2 + {\Phi \over r} \longrightarrow + \infty,   $$
and this contradicts (5.3).    []

Note that Lemmas 5.14, and 5.15 prove Proposition 5.10.

\noindent
\underline{Corollary 5.16}.  {\it If (5.2) and (5.3) hold, then $w(r)$ is of 
one sign for $r$ near $r_0$.}

We next show that for $r$ near $r_0$,
\begin{equation}
{\rm either} \ \ \ w^2(r) > 1 \ \ \ {\rm or} \ \ \ w^2(r) < 1;
\end{equation}
that is, either $w < -1$, or $-1 < w < 0$, or $0 < w < 1$, or $w > 1$.  To
 prove this we need two lemmas, the first of which is:

\noindent
\underline{Lemma 5.17}.  {\it If (5.2) and (5.3) hold then $lim_{r \searrow r_0}\; w^2(r)=1$ is not possible.}

\noindent
\underline{Proof}.  Suppose (for definiteness) that $lim_{r \searrow r_0}\; w(r)= -1$.  With $\ve$ defined by (5.30), we see that for $r$ near $r_0$, $-1 -\ve \le w(r) \le -1 + \ve$. On this interval, (5.31) implies $\Phi(r) > .9r_0$.  Then from (2.6), 
$$	rA' = -2Aw'^2 + {\Phi \over r} > {.9r_0 \over r} > 0,    $$
and this contradicts (5.3). \ \ \ \ []

\medskip

We next show that the orbit has finite rotation about $(1,0)$ in the case $w > 0$ near $r_0$, or about $(-1,0)$ in case $w < 0$.

\noindent
\underline{Lemma 5.18}.  {\it If (5.2) and (5.3) hold and $w > 0$ for $r$ near $r_0$, then the projection of the orbit in the $w-w'$ plane has finite rotation about $(1,0)$.  Similarly if $w < 0$ for $r$ near $r_0$, then the projection of the orbit in the

 $w-w'$ plane has finite rotation about $(-1, 0)$.} 

\noindent
\underline{Proof.}  Suppose $w > 0$ near $r_0$ (the proof for $w < 0$ is similar, and will be omitted), and the orbit has infinite rotation about (1,0).  Since $\lim_{r \searrow r_0} w(r)  \not= 1$, we must have either
$ \overline{\lim}_{r\searrow r_0} w(r) > 1$ or  $\overline{\lim}_{r\searrow r_0} w(r) < 1$.
In either case we repeat the argument of Lemma 5.10 using the $w$-interval 
$[1, 1+\ve]$ or $[1 - \ve, 1]$.  We have that $\Phi$ is bounded away 
from $0$ by (5.31).  By Lemma 5.13, either $(Aw'^2)(r) \ra -\infty$ as $r 
\searrow r_0$, or $Aw'^2$ is bounded 
near $r_0$.  We rule out the case $Aw'^2 \ra -\infty$ because $w'$ is of one 
sign; hence $Aw'^2$ is bounded near $r_0$.  Using Theorem 2.3 exactly as in 
Lemma 5.14, we have that the orbit can cross the line $w=1$ a finite number
 of times.  Thus $w > 1$ or
$w < 1$ for $r$ near $r_0$.     []

Summarizing, we have

\noindent
\underline{Corollary 5.19}.  {\it For $r$ near $r_0$, precisely one of 
the following holds: $w(r) < -1, \ -1  < w(r) < 0, \ 0 < w(r) < 1$, or 
$w(r) > 1$.}

\smallskip

Since $w''$, when $w' = 0$, has a fixed sign in each of the four strips, we 
see that $w'$ must have a fixed sign for $r$ for $r_0$; i.e., the projection
 of the orbit in the $w - w'$ plane must lie in one of the 8 regions depicted 
 in Figure 7.  Since we now have the orbit confined to one of these 8 
 regions, without loss of generality we will consider the case where $w' < 0$.
   
We will first show that orbit cannot lie in regions (6) or (8) for $r$ 
near $r_0$.  Then we will show that if the orbit is in regions (5) or (7), 
and $w'$ is bounded near $r_0$, that $\lim_{r \searrow r_0} A(r) = 0$ and
 $\lim_{r \searrow r_0} (w(r), w'(r))$ exists and lies on $C_{r_0}$; 
 hence the orbit continues past $r_0$.  We complete the proof of Theorem 5.1 
 by showing that the case where $w'$ is unbounded near $r_0$ cannot occur.

\smallskip
\noindent
\underline{Lemma 5.20}.  {\it If (5.2) and (5.3) hold, then the orbit 
cannot lie in regions (6), or (8) for $r$ near $r_0$.}

\noindent
\underline{Proof.}  In regions (6) and (8), $w$ is bounded near $r_0$.  Thus from Lemma 5.11,
\begin{equation}
\lim_{r \searrow r_0} A(r) = 0.
\end{equation}
\vfill\eject
If $v = Aw'$, then from (2.13) we see $v' \le 0$ so $\lim_{r \searrow r_0} v(r) = L > 0$ exists.  Thus writing $Aw'^2 = {v^2 \over A}$, we see that 
\begin{equation}
\lim_{r \searrow r_0} (Aw'^2)(r) = -\infty.
\end{equation}
Since $w$ is bounded near $r_0$ (5.33) implies that $\Phi$ is bounded near $r_0$.  Thus, from (2.6),
$$    rA' = {\Phi \over r} -2Aw'^2 \longrightarrow +\infty     $$
as $r \searrow r_0$.  However, this contradicts (5.3). \ \ \ []

\medskip

We now consider the case where (5.2) and (5.3) hold, and the orbit lies 
in one of the regions (5) or (7) for $r$ near $r_0$,\  $r > r_0$.

We first consider the case where $w'$ is bounded.

\smallskip
\noindent
\underline{Lemma 5.21}.  {\it Suppose that (5.2) and (5.3) hold, and that the orbit lies in either region (5) or (7) for $r$ near $r_0$.  If $w'(r)$ is bounded near $r_0$ then 
$\lim_{r \searrow r_0} A(r) = 0$, $\lim_{r \searrow r_0}$ $ (w(r),
 w'(r)) = (\bar{w}, \bar{w}')$ exists, and $(\bar{w}, \bar{w}')$ lies 
 on $C_{r_0}$. }

Note that in view of our remark preceding Proposition 5.10, Lemma 5.21 implies 
that Theorem 5.1 holds in this case.

\noindent
\underline{Proof.}  First note that since $w'$ is bounded, this implies $w$ is bounded, and hence Lemma 5.11 implies that
\begin{equation}
\lim_{r \searrow r_0} A(r) = 0.
\end{equation}
Now as $A \ra 0$, and $w$ has a limit, we see that $\Phi = r - rA - u^2/r$ has a limit; call this limit $\Phi_0$; i.e.
\begin{equation}
\Phi_0 = \lim_{r \searrow r_0} = \Phi(r).
\end{equation}
If $\Phi_0 \not= 0$, then as $\lim_{r \searrow r_0} v(r) = 0$  we may apply L'Hospital's rule to obtain
\begin{eqnarray*}
\lim_{r \searrow r_0} w'(r) &=& \lim_{r \searrow r_0} \  {v(r) \over A(r)} = \lim_{r \searrow r_0} {v'(r) \over A'(r)}  \\
 \\
&=& \lim_{r \searrow r_0} \  {{-2w'^2v \over r} - {uw \over r^2} \over {\Phi \over r^2} - {2Aw'^2 \over r}} = \lim_{r \searrow r_0} \ \left[ {-uw \over \Phi} \right],
\end{eqnarray*}
where we have used (2.6) and (2.13).  Thus
\begin{equation}
\lim_{r \searrow r_0}w'(r) = lim_{r \searrow r_0} \  \ \left[ {-uw \over \Phi} \right].
\end{equation}
We claim that
\begin{equation}
\Phi_0 \not= 0.
\end{equation}
Note that if (5.38) holds, then since $w$ has a finite limit at $r_0$, (5.37) implies that $\lim_{r \searrow r_0} w'(r)$ exists and is finite, and
$$  \lim_{r \searrow r_0} (w(r), w'(r)) \in C_{r_0}.  $$
Thus, to complete the proof Lemma 5.21, it suffices to prove (5.38).

Thus, assume $\Phi_0 = 0$; we show this leads to a contradiction.  If $(uw)(r_0) \not= 0$, then (5.37) implies that $w'(r)$ is unbounded near $r_0$, and this is a contradiction.  Hence we may assume $(uw)(r_0) = 0$.  If $u(r_0) = 0$, then
$$  0 = \Phi_0 = r_0 - {u^2_0 \over r_0} = r_0 \; , $$
and this is a contradiction since $r_0 > 0$.  Thus we may assume $w(r_0) = 0$.  In this case 
$$     0 = \Phi_0 = r_0 - {1 \over r_0} \ , $$
so that   $$  r_0 = 1.$$
Note too that if $w(r_0) = 0$, the orbit lies in region (7) for $r$ near $r_0$.  We now have
\begin{equation}
A(r_{n+1}) - A(r_n) = (r_{n+1} - r_n)A'(\xi)
\end{equation}
where $r_n > \xi > r_{n+1} > 1$.  From (2.6)
\begin{equation}
\xi A'(\xi) = 1 - A(\xi) - {u^2(\xi) \over \xi^2} - 2(Aw'^2)(\xi).
\end{equation}
Since $\xi > 1, \ 1 - {u^2(\xi) \over \xi} > 0$, so for large $n$, (5.40)
 implies $A'(\xi) > 0$.  Using this in (5.39) gives $0 > A(r_n) > A(r_{n+1})$, 
 and this violates (4.3).  Thus (5.38) holds and the proof is complete. 
 \ \ \ []

We now consider the case where (5.2) and (5.3) hold, and the orbit is in 
region (5) or (7), and $w'(r)$ is unbounded for $r$ near $r_0$, $r > r_0$. 
 We shall show that this case is impossible.

First note that if $w$ is bounded near $r_0$, it follows from Lemma 5.11,  
that
\begin{equation}
\lim_{r \searrow r_0} A(r) = 0.
\end{equation}
Since $w' < 0$, $\lim_{r \searrow r_0} w(r)$ exists. Thus if $w$ is bounded near $r_0$, $\lim_{r \searrow r_0} \Phi(r)$ exists and is finite; say
\begin{equation}
\lim_{r \searrow r_0} \Phi(r) = \Phi_0.
\end{equation}
We now have

\smallskip
\noindent
\underline{Proposition 5.22}.  {\it If (5.2) and (5.3) hold, and $w'$ is unbounded near $r_0$,  then $w$ cannot be bounded near $r_0$; in particular that orbit cannot lie in region (7).}

\noindent
\underline{Proof.}  Suppose that $w(r)$ is bounded for $r$ near $r_0$; we 
will show that this leads to a contradiction. 

Thus, in this case (5.41) holds and $\Phi_0$ is finite.  We consider 3 
cases $\Phi_0 > 0$, $\Phi_0 < 0, \ \Phi_0 = 0$, and we will obtain 
contradictions in all cases.

\noindent
\underline{Case 1.} $\Phi_0 > 0$.

From (2.6), for $r$ near $r_0$
$$    A'(r) = {\Phi \over r^2} - {2Aw'^2 \over r} > 0,    $$
and this violates (5.3); thus Case 1 cannot occur.

\noindent
\underline{Case 2.}  $\Phi_0 < 0$.

We first show
\begin{equation}
\lim_{r \searrow r_0} w'(r) = -\infty.
\end{equation}
To see this, note that if (5.43) were false, then as $w'$ is unbounded near $r_0$, there would exist a sequence $s_n \searrow r_0$ such that
$$	w'(s_n) < -n \ \ \ {\rm and} \ \ \ w''(s_n) = 0 .      $$
Then from (2.7)
\begin{eqnarray*}
0 &=& s^2_n(Aw'')(s_n) + \Phi(s_n)w'(s_n) + (uw)(s_n) \\
&=& \Phi(s_n) w'(s_n) + (uw)(s_n) \longrightarrow \infty 
\end{eqnarray*}
as $n \ra \infty$.  This contradiction implies that (5.43) holds.

Now if $f = Aw'^2$, then from (2.14)
\begin{equation}
r^2f' + (2rf + \Phi)w'^2 + 2uww' = 0,
\end{equation}
and since $(2rf + \Phi)$ is strictly negative near $r_0$ and $w$ is bounded near $r_0$ it follows from (5. 43) that $f'(r) > 0$ if $r$ is near $r_0)$.  Thus
$$    \lim_{r \searrow r_0}  f(r) = L < 0   $$
exists; where $L \ge -\infty.$  We claim that
\begin{equation}
L = - \infty.
\end{equation}
To see this, we note first that
\begin{equation}
(w'^2v)(r) = w'(r)f(r) \ra + \infty, \ \ \ (v = Aw'),
\end{equation}
so that (c.f. (2.13)),
$$     v' = {-2w'^2v \over r} - {uw \over r^2} \ra -\infty,    $$
since $w$ is bounded near $r_0$.  Hence, if $r_0 < r < r_1$, and $r_1$ is near $r_0, \ v(r_1) < v(r)$ so
$$     (Aw'^2)(r) = v(r)w'(r) < v(r_1)w'(r) ,$$
and as $v(r_1) w'(r) \ra - \infty$, we see that $(Aw'^2) (r) \ra -\infty$.  As $r \searrow r_0$; thus (5.45) holds.

Now again using (2.6),
$$     rA'(r) = -2(Aw'^2)(r) + {\Phi \over r} \ra +\infty,  $$
as $r \searrow r_0$.  But this violates (5.3); hence Case 2 cannot occur.  
We now turn to the final case,

\smallskip
\underline{Case 3.}  $\Phi_0 = 0$.

The proof in this case relies on Theorem 2.2.  Indeed, we will show that 
$\lim_{r\searrow r_0} A'(r) = 0$, and from (5.41), $\lim_{r\searrow r_0} 
A(r) = 0$,   This is enough to invoke Theorem 2.2, to conclude that $w(r) 
\equiv 0$ and thus $w'(r) \equiv 0$; this violates the assumption that $w'$ 
is unbounded.

We first show
\begin{equation}
\lim_{\overline{r \searrow r_0}} A'(r) \le 0.
\end{equation}
Indeed, if $\underline{\lim}_{r \searrow r_0} A'(r) > 0$ then for $r > r_0$, $r$ near $r_0$,
$$    0 > A(r) = A(r) - A(r_0) = (r-r_0)A'(\xi) > 0   $$
where $\xi$ is an intermediate point.  This contradiction establishes (5.47).  

Next, since
\begin{equation}
rA' = {\Phi \over r} - 2Aw'^2 \ ,
\end{equation}
it follows from (5.47) that $\underline{\lim}_{r \searrow r_0} ({\Phi \over r} - 2Aw'^2) \le 0$, so $\underline{\lim}_{r \searrow r_0}$  $({\Phi_0 \over r_0} - 2Aw'^2) \le 0$, or
$$    0 \ge \overline{\lim}_{r \searrow r_0} 2Aw'^2 \ge {\Phi_0 \over r_0} = 0  $$
thus
\begin{equation}
\overline{\lim}_{r \searrow r_0} \  Aw'^2 = 0.
\end{equation}

We next show
\begin{equation}
\lim_{\overline{r\searrow r_0}} Aw'^2 = \overline{\lim}_{r \searrow r_0} \  Aw'^2.
\end{equation}
(Note that if (5.50) holds, then $\lim_{r \searrow r_0} Aw'^2 = 0$, so from (5.48) $A'(r_0) = 0$.   Thus the proof of Proposition 5.22 will be complete once we prove (5.50).)

So suppose that there is an $\eta > 0$ such that
\begin{equation}
\lim_{\overline{r\searrow r_0}} Aw'^2 \le -2\eta \ .
\end{equation}
Then in view of (5.49), if $f = Aw'^2$, we can find a sequence $s_n \searrow r_0$ such that $f(s_n) = -\eta, \ f'(s_n) < 0$.  Since (5.41) holds, we have $A(s_n) \ra 0$ so that
$w'(s_n) \ra -\infty$.  From (5.44),
\begin{equation}
s^2 f'(s_n) + (-2s_n \eta + \Phi(s_n))w'^2(s_n) + 2(uww')(s_n) = 0.
\end{equation}
But as $f'(s_n) < 0$ and $w'^2(s_n) \ra \infty$, we see that (5.52) cannot hold for large $n$.  Thus (5.50) holds and this implies $\lim_{r \searrow r_0} Aw'(r) = 0$, and thus by Theorem 2.2, we have a contradiction. \ \ \ \ []

\medskip

We now consider the final case in the proof of Theorem 5.1, namely in regions (5) or (7)
\begin{equation}
w \ {\rm and} \ w' \ {\rm unbounded \ near} \ r_0.
\end{equation}
(Of course, this implies that we are in region (5)).  Note too that in this case we have
\begin{equation}
\lim_{r \searrow r_0} w(r) = + \infty.
\end{equation}

\noindent
\underline{Proposition 5.23}. {\it If (2.2) and (2.3) hold, and the 
orbit lies in region (5), then (5.54) cannot hold.}

Note that once Proposition 5.23 is established this will complete 
the proof of Theorem 5.1.

\noindent
\underline{Proof.}  From our remark following the statement of Lemma 5.6, we have
\begin{equation}
\lim_{r \searrow r_0} w'(r) = -\infty.
\end{equation}
Then as we have remarked earlier (5.18) holds; i.e. $Aw'^2 > w^5$, for $r$ near $r_0$.  Thus, from (5.48) for $r$ near $r_0$,
$$  rA'(r) = -2f + {\Phi \over r} > 2w(r)^5 + \left( 1 - A - {u^2 \over r} \right) > 0,   $$
since $u^2$ is of order $w^4$, and this contradicts (5.3). \ \ \ \ []

\vfill\eject

\setcounter{section}{5}
\section{Miscellaneous results and open questions.}
\setcounter{equation}{0}

\ \ \ \ \ In Section 3, we proved that the zeros of $A$ are discrete, except 
possibly at $r=0$.  This leads to the first question.

\smallskip

1.  Can $r=0$ be a limit point of zeros of $A$?

\smallskip
\noindent
We conjecture that the answer is no.  In a recent paper [4, p. 8, $\ell$ 7], 
the authors assume that the answer is no.  A rigorous proof of this would be 
welcome.

A related question is

\smallskip

2.  Do there exist solutions of the EYM equations for which $A$ has more than 
two zeros?

\smallskip
\noindent
A negative answer obviously implies a negative answer to Question 1.  In [5],
 the authors have numerically obtained a solution having two zeros.  This 
 leads to the next Problem.

3.  Give a rigorous proof of the existence of a global solution of the 
EYM equations, (other than the classical Reissner-Nordstr\"{o}m solution), 
where $A$ has two zeros.

\smallskip

4.  A subject of much current interest is the study of solutions near $r=0$; 
[4,5].  If, as we suspect, Question 1 has an affirmative answer, then every
 solution near $r=0$, has either $A > 0$ or $A < 0$.  If $A > 0$ near $r=0$, 
 then we have proved in [10], that either $\lim_{r \searrow 0} A(r) = 1$,
  in which case the solution is particle-like, or else $\lim_{r \searrow 0} 
  A(r) = +\infty$, in which case the solution is a Reissner-Nordstr\"{o}m-Like 
  (RNL) solution, [14]; this case is re-discussed in [4].

If $A < 0$ near $r=0$, much less is known.  In [14], we proved the following 
theorem:

\smallskip
\noindent
\underline{Theorem 6.1}. {\it Given any triple of the form $q=(1,b,c)$, 
there exists a unique local RNL solution $(A_q(r), w_q(r))$, 
satisfying $\lim_{r \searrow 0} rA(r) = b$, $w_q(0) = 1$, $w''_q(0) = c$, 
and the solution depends continuously on these values.}

\smallskip

If $b < 0$, then $\lim_{r \searrow 0} A(r) = -\infty$, and $\lim_{r 
\searrow 0} (w^2(r), w'(r)) = (1,0)$.  These solutions have been termed
 Schwarzschild-like [5].  In [5], the authors also investigated RNL 
 solutions but they mistakenly omitted the 2-parameter family of solutions 
 that have $w(0) = 0$.  These solutions have the following asymptotic form
  near $r=0:$
\begin{eqnarray*}
A(r) &=& {1 \over r^2} + {b \over r} + h.o.t. \\
w(r) &=& cr^3 + h.o.t.
\end{eqnarray*}
These solutions are interesting since they give rise to asymptotically flat 
solutions with half-integral rotation numbers; see [14].  In addition the 
authors of [5] omitted solutions which have $w^2(0) = 1$; these solutions 
have the following asymptotic form near $r=0:$
\begin{eqnarray*}
A(r) &=& {b \over r} + h.o.t. \\
w(r) &=& \pm 1 + cr^2 + h.o.t.
\end{eqnarray*}
There is still another type of local solution, (discussed in [5]), having $A < 0$ near $r=0$, but these do not appear to give rise to asymptotically flat global solutions, [5].  We are thus lead to the following `trichotomy conjecture":

\smallskip
\underline{Conjecture}:  If $(A(r), w(r))$ is a globally defined solution of the EYM equations (1.3), (1.4), then
$$
\lim_{r \searrow 0} A(r) = \left\{ 
\begin{array}{cc}
-\infty, \ & \ {\rm or} \\
+1, \ & \ {\rm or} \\
+ \infty .&
\end{array}
\right.
$$
In view of our above remarks concerning the behavior of solutions if $A(r) > 0$ near $r=0$, this conjecture can be rephrased as:

\smallskip
\noindent
\underline{Conjecture}: If $(A(r), w(r))$ is a globally defined solution to the EYM equations (1.3), (1.4), and $A(r) < 0$ for $r$ near $0$, then $\lim_{r \searrow 0} A(r) = -\infty$

\vfill\eject

5.  Another interesting question is the following:

Does there exist a solution to the EYM equations (1.3), (1.4), where 
$A(r) < 0$ in a neighborhood of $r=\infty$?

\smallskip

We conjecture that the answer to this question is negative.  If our 
conjecture is true, this would enable us to drop the hypothesis 
$A(\bar{r}) > 0$ in Theorem 1.2.  If, on the other hand the 
conjecture is true, then we can show that the orbit must have infinite
 rotation in the $(w,w')$-plane and $w$ must be unbounded.

\smallskip

6.  Using the methods in [7, 8, 9], we have proved the following theorem

\smallskip
\noindent
\underline{Theorem 6.2}. {\it There is a continuous 2-parameter family of 
solutions $\left(A_{\alpha,\beta}(r), w_{\alpha,\beta}(r)\right)$ to the 
EYM equations (1.3), (1.4), defined in the far-field, which are analytic 
functions of $s = {1 \over r}$.  That is, if $(A(r), w(r))$ is a solution to 
the EYM equations (1.3), (1.4) which is asymptotically flat, and is analytic 
in $s = {1 \over r}$, then $(A(r),w(r))$ = $\left(A_{\alpha,\beta}(r), 
w_{\alpha,\beta}(r)\right)$ for some pair of parameter values $(\alpha,\beta)$.}

\noindent
(We omit the details of the proof as they are similar to those in [7].)

\smallskip

In the above theorem, one parameter is the (ADM) mass $\beta$, and in fact,
 $A(s=0) = 1$, and ${dA \over ds}\big|_{s=0} = -\beta$.  The other parameter 
 is $\alpha = {dw \over ds}\big|_{s=0}$, and $w^2(s=0) = 1$; c.f. [10].

It follows from the results in [10 or 14], that the (ADM) mass $\beta$ is 
finite for any solution which is defined in the far-field.  Moreover, for
 such solutions $\lim_{r\ra \infty} rw'(r) = 0$; c.f. [9].  We do not know 
 whether $\lim_{r \ra \infty} r^2 w'(r) \equiv \lim_{r \ra 0} {dw(s)
  \over ds}$ exists.  This leads to the next question:

Is every asymptotically flat solution to the EYM equations (1.3), (1.4) 
analytic in $s = {1 \over r}$ at $s=0$?

If the answer is affirmative, then we may consider the $(\alpha,\beta)$-plane 
as representing those solutions having the following asymptotic form near
 $s=0$:
\vfill\eject
\begin{eqnarray*}
A(s) &=& 1 - \beta s + h.o.t. \\
w(s) &=& 1 - \alpha s + h.o.t. ,
\end{eqnarray*}
and all such solutions are described by a point in the $(\alpha,\beta)$-plane,
 (or in the corresponding plane corresponding to $w(s=0) = -1)$, or they
  correspond to the 1-parameter family of classical 
  Reissner-Nordstr\"{o}m solutions:  $A(r) = 1 - {c\over
r} + {1 \over r^2}, \ w(r) \equiv 0$.

We consider the $(\alpha, \beta)$-plane as depicted in Figure 8.  
In this plane, certain regions 
are easy to identify.  Thus, if $\alpha < 0$, these correspond to 
RNL-solutions.  Similarly, the region $\alpha > 0, \beta < 0$,
 also correspond to RNL-solutions.  The line $\alpha = 0$ corresponds to 
 Schwarzschild solutions with mass $\beta$.  Particle-like and black-hole 
 solutions must lie in the 1st quadrant $\alpha > 0, \ \beta > 0$. 
  Presumably, there are a countable number of curves in the 1st quadrant 
  distinguished by the number of zeros of $w$, parametrized by $\rho$, the 
  event horizon.  (These are schematically depicted in Figure 8, where the 
  points $P_n$ correspond to particle-like solutions and the $\beta$ 
  coordinate of $P_n$ tends to $2$ as $n \ra \infty$; c.f. [11].)  There 
  are also a countable number of points in this quadrant which correspond to
   particle-like solutions. 

Thus, near any particular black-hole solution, there are global solutions 
which are neither black-hole or particle-like solutions; i.e., they must be 
RNL solutions.  This follows since any point in this plane represents a 
global solution (from our results
in this paper; c.f. Theorem 1.2).  Thus for any such global solution $(A,w)$,
 either $A$ has a zero, in which case the corresponding point 
 $(\alpha,\beta)$ lies on one of the above-mentioned countable number 
 of curves, or it is one of the countable number of particle-like solutions, 
 or it is an RNL solution [10, 14].

It follows that in any neighborhood of a black-hole solution $(A_0(r),
 w_0(r))$ there are RNL solutions.  In particular, if $A_0(r_1) = -\eta < 0$, 
 then arbitrarily close to this solution, there are solutions $(A(r), w(r))$ 
 having $A(r_1) > 0$.  This is a spectacular example of non-continuous 
 dependence on initial conditions.

\vfill\eject

\centerline{REFERENCES}

\begin{enumerate}
\item Bartnik. R., and McKinnon, J., ``Particle-like solutions of the Einstein-Yang-Mills equations", {\it Phys. Rev. Lett.} 61, 141-144, (1988).
\item Bizon, P., ``Colored black holes", {\it Phys. Rev. Lett. 64}, 2844-2847, (1990).
\item Breitenlohner, P., Forg\'{a}cs, P. and Maison, D. ``Static spherically symmetric solutions of the Einstein-Yang-Mills equations", {\it  Comm. Math. Phys.} 163, 141-172, (1994).
\item Breitenlohner, P.,  Forg\'{a}cs, P. and Maison, D., ``Mass inflation and chaotic behavior inside hairy black holes", gr-qc/9703047.
\item Donats, E.E., and Gal'tsov, D.V., ``Internal structure of Einstein-Yang-Mills black holes", gr-qc/9612067.
\item Kunzle, H.P., and Masood-ul-Alam, A.K.M., ``Spherically symmstric static SU(2) Einstein-Yang-Mills fields", {\it  J. Math. Phys.} 31, 928-935 (1990).
\item Smoller, J., Wasserman, A., Yau, S.-T., McLeod, J., ``Smooth static solutions of the Einstein-Yang Mills equations", {\it  Comm. Math. Phys.} 143, 115-147 (1991).
\item Smoller, J., Wasserman, A., ``Existence of infinitely-many smooth static, global solutions of the Einstein/Yang-Mills equations", {\it  Comm. Math. Phys.} 151, 303-325, (1993).
\item Smoller, J., Wasserman, A., and Yau, S.-T., ``Existence of black hole solutions for the Einstein-Yang/Mills equtions", {\it  Comm. Math. Phys.} 154, 377-401, (1993).
\item Smoller,J., and Wasserman, A., ``Regular solutions of the Einstein-Yang/Mills equations", {\it  J. Math. Phys.} 36, 4301-4323, (1995).
\item Smoller, J. and Wasserman, A., ``Limiting masses of solutions of Einstein-Yang/Mills equations, {\it  Physica D.}, 93, 123-136, (1996).
\item Smoller, J. and Wasserman, A., ``Uniqueness of extreme Reissner-Nordstr\"{o}m solution in SU(2) Einstein-Yang/Mills theory for spherically symmetric spacetime", {\it  Phys. Rev. D.}, 15 Nov. 1995., 52, 5812-5815, (1995).
\item Smoller, J., and Wasserman, A., ``Uniqueness of zero surface gravity SU(2) Einstein-Yang/Mills black holes", {\it  J. Math. Phys.} 37, 1461-1484, (1996).
\item Smoller, J., and Wasserman, A., ``Reissner-Nordstr\"{o}m-Like solutions of the SU(2) Einstein-Yang/Mills equations", (preprint, see announcement gr-qc/9703062).
\item Straumann, N., and Zhou, Z., ``Instability of a colored black hole solution", {\it  Phys. Lett B.} 243, 33-35, (1990).
\item Ershov, A.A., and Galtsov, D.V. ``Non abelian baldness of colored black holes, {\it  Physics Lett. A.} 150, 747, 160-164 (1989).
\item Lavrelrashvili, G., and Maison, D., ``Regular and black-hole solutions of Einstein-Yang/Mills dilation theory", {\it  Phys. Lett. B.} 295, (1992), 67.
\item Volkov, M.S. and Gal'tsov, D.V., ``Black holes in Einstein-Yang/Mills theory", {\it  Sov. J. Nucl. Phys.} 51, (1990), 1171.
\item Volkov, M.S., and Ga.'tsov, D.V., ``Sphalerons in Einstein-Yang/Mills theory", {\it  Phys. Lett. B.} 273,  (1991), 273.
\end{enumerate}

\vspace{.25in}

\noindent
University of Michigan \\
\noindent
Mathematics Department \\
\noindent
Ann Arbor, MI  48109-1109\\
USA

\end{document}